\documentclass[12pt, a4paper, makeidx]{article}
\usepackage {amssymb,amsmath}
\usepackage{graphicx}
\usepackage{float}
\usepackage[top=0.8in, bottom=0.8in, left=0.8in, right=0.8in]{geometry} 

\begin{document}
\hyphenchar\font=-1
\sloppy

\begin{Large}
\begin{center}
\textbf{Sausage Mode Propagation in a Thick Magnetic Flux Tube}
\end{center}  
\end{Large}
\vspace{40pt}

\begin{flushleft}
\textbf{A. Pardi, I. Ballai, A. Marcu, B. Orza}

\end{flushleft}

\section*{Abstract}
The aim of this paper is to model the propagation of slow
magnetohydrodynamic (MHD) sausage waves in a thick expanding
magnetic flux tube in the context of the quiescent (VAL C) solar
atmosphere. The propagation of these waves is found to be
described by the Klein-Gordon equation. Using the governing MHD
equations and the VAL C atmosphere model we study the variation of
the cut-off frequency along and across the magnetic tube guiding the waves. Due to the radial
variation of the cut-off frequency the flux tubes act as low
frequency filters for waves.

Keywords: Magnetohydrodynamics (MHD), Sun, Waves

\section*{Introduction}
The dynamical response of the plasma in the solar atmosphere to
rapid changes in the solar interior can be manifested through
waves propagating within the solar atmosphere. The dynamics of a
large class of waves with wavelengths and periods large compared
with the ion Larmor radius ($10^{-4}$ m in the solar photosphere
and about 1 m for almost all combinations of coronal parameters)
and the gyroperiod ($10^{-7}$ s in the solar photosphere and of
the order of $10^{-3}$ s in the solar corona), respectively, can
be described within the framework of collisional
magnetohydrodynamics. Waves perturb macro-parameters of the
plasma, such as density, temperature, bulk velocity and magnetic
field. The period range of waves and oscillations observed in the
solar atmosphere (from a few seconds to a few hundreds of minutes)
is well covered by temporal resolution of presently available
ground-based and spaceborne observational telescopes. The study of
waves and oscillations in the context of solar atmospheric physics
has a long history that started more than six decades ago. While
until very recently this study was driven mainly by scientific
interest, the continuous development and refinement of this
important branch of solar and space physics was motivated by the
tremendous amount of observations showing the propagation
characteristics of waves and oscillations in various solar
structures. Nowadays almost all waves labelled as {\it
magnetohydrodynamic waves and oscillations} are observed with
great accuracy. In this context slow MHD waves were observed in
the solar photosphere (Zirin72, Dorotovic08, Fujimura09), in
the solar chromosphere e.g. Gilbert04, solar corona e.g.
Schrijver1999, Nightingale99a, Nightingale99b, Berghmans99, De
Moortel00, McEwan06, Berghmans01, Erd08, Marsh09 and solar plumes
e.g. DeForest98, Ofman2000. Although initially their
observations was not easy (due to their very short period),
several evidences were adduced to prove the existence of fast
waves e.g.
Asai01, Williams01, Williams02, Katsiyannis03, Dorotovic12, in
different magnetic structures. A special note should be made
regarding the large number of fast kink oscillations of coronal loops
observed recently e.g.
Aschwanden99, Wang03, Ofman2008, Ballai2011 that served as
key ingredient in the development of coronal seismology.

Until recently wave propagation in magnetic guides were modelled
especially in the thin flux tube limit (for a review see, e.g.
Roberts2001, supposing that the wave length of waves is
much larger than
 the radius of the tube (i.e. $kR\ll1$). This approximation allows modelling the propagation of waves and
 oscillations in various magnetic structures in a tractable way since the radial dependence of perturbations
  is factorised out. While this is a viable limit for waves and oscillations in the solar corona and upper part
   of the chromosphere where wave lengths are comparable to the length of loops RaeRoberts1982, in the
   solar photosphere this condition is hardly satisfied. Here, the thin flux tube approximation can be applied
   to only very thin magnetic structures often found at the edges of granules. In general the structure of the
   waveguide for waves propagating in the solar photosphere the tube can be considered thick. It is well known
   that under photospheric conditions the density scale height is of the order of a few hundred kilometers, so any wave
   that has a wavelength equal or larger than this length will experience the effect of gravitational stratification.
    If we assume that the wavelength is comparable to the scale-height, the condition of applicability of the thin
    flux tube approximation reduces to $2\pi R/H\ll 1$. Assuming an isothermal atmosphere the exact value of the scale-height
     can be determined from the VAL-C model (Vernazza1981). This criterion will be addressed later in our investigation.

In our study we investigate the propagation of waves in magnetic structures that ascend from the convective zone,
open up as they reach the chromosphere and then either continue in the corona as opened magnetic flux tubes (Fedun2011)
or become horizontal and eventually descend, forming the magnetic canopy.

Magnetic structures channel the propagation (among other various types of waves)
 of slow longitudinal acoustic waves (for a detailed review see e.g. Roberts2006),
 and the propagation of these waves in the presence of gravity in the thin flux tube approximation is described by the Klein-Gordon
 equation (see also  Ballai2006, Erdi108). The solution of  the Klein-Gordon equation in this case
 describes the behaviour of slow MHD waves propagating with a phase speed close to the tube speed and the propagation of a wake
 following the wave that propagates with the cut-off frequency. The cut-off frequency in this context is defined as a threshold
 limit of frequencies and represents the lower value of frequencies describing the propagating waves. In the opposite case, when the frequencies
 of waves are smaller than the cut-off frequency, waves will become evanescent, i.e. eigenfunctions will fall-off exponentially with distance.
 Observations show (since 1964)  (Aschwanden2006) that the cut-off frequencies for these longitudinal axi-symmetric displacements vary on
 average between 0.025 and 0.03 Hz.

The paper is organised as follows: in Section 2 we will
describe the thick expanding magnetic flux tube model (and obtain
the equations of the magnetic field) and we will determine the
evolutionary equation of slow sausage waves propagating in
structures of this type. Using simple mathematical methods and
straightforward assumptions we will reduce the evolutionary
equation to a Klein-Gordon type equation and we will determine the
expression of the cut-off frequency for thick magnetic tubes in
Section 3. In Section 4 the formulaic
expression will be joined with a realistic atmospheric model that
will help us study the evolution of the cut-off frequency in
longitudinal and transversal direction of the tube. Finally our
results will be summarised and conclusions will be drawn in
Section 5.

\section*{The mathematical and physical considerations of the problem }

We consider a gravitationally structured elastic magnetic flux
tube that at the height $z=0$ has a radius $R_{0}$ and a cross section
$A_{0}$. Due to decrease of the equilibrium parameters (density
and pressure) with height, the tube is expanding in the horizontal
direction. We will consider a 2D equilibrium magnetic field with
components $\textbf{B}_{0}=(B_{r},0,B_{z})$, each one of them
being both \emph{r} and \emph{z} dependent. For simplicity we
assume that the equilibrium magnetic field is potential (a special
case of force-free fields). As a result, the equations that can
fully describe the structure of the magnetic field are
\begin{equation}
\nabla \times {\bf B_0}=0, \quad \nabla\cdot {\bf B_0}=0,
\label{eq:1.1}
\end{equation}
where the second equation is the solenoidal condition.

Written in cylindrical geometry, the above two equations can be reduced to a PDE for the $z$-component of the magnetic field
\begin{equation}
\frac{1}{r}\frac{\partial B_{z}}{\partial r}+\frac{\partial^{2}
B_{z}}{\partial r^{2}}+\frac{\partial^{2} B_{z}}{\partial z^{2}}=0.
\label{40}
\end{equation}

Equation (\ref{40}) can be solved by using the separation of the
variables method in which case we write $B_{z}=F_{1}(r)F_{2}(z)$
therefore Eq.(\ref{40}) becomes
\begin{equation}
\frac{F_{1}^{'}(r)}{F_{1}(r)r}+\frac{F_{1}(r)^{''}}{F_{1}(r)}=-\frac{\ddot{F_{2}}(z)}{F_{2}(z)}=\lambda,
\label{41}
\end{equation}
where $\lambda$ is a separation constant, the {\it dash} and {\it dot} represent differentiation with respect to $r$ and $z$, respectively. It can be easily shown
that the only physically acceptable case is when
$\lambda=-\nu^2<0$. In this case the solutions of Eq. (\ref{41})
become
\begin{equation}
F_{1}(r)=d_{1}J_{0}(\nu r)+d_{2}Y_{0}(\nu r), \label{43}
\end{equation}
\begin{equation}
F_{2}(z)=d_{3}e^{\nu z}+d_{4}e^{-\nu z}, \label{44}
\end{equation}
where  $J_{0}$ and $Y_{0}$ are zeroth order Bessel functions and the quantities $d_i$ ($i=1-4$) are constants that can be
determined when considering the suitable boundary conditions.

Our intention is to construct such a magnetic field model which that at $z=0$ and at the
 centre of the tube has perfectly vertical field lines, i.e.
  $B_{z}(r=0,z=0)=B_{0}, B_{r}(r=0,z=0)=0 $.  As we go higher up in height and further
 away from the axis, the field lines become more and more tilted with respect to
 the vertical direction and eventually they become horizontal far away from the
 vertical symmetry axis. Since the function $Y_{0}(\nu r)$ is divergent at $r=0$ we
 will choose $d_{2}\equiv0$, so the radial component reduces to $F_{1}(r)=d_{1}J_{0}(\nu r)$. In addition since observations show that the cross-section of the tube becomes wider (in virtue of the equation of flux conservation) with height we expect that the $z$-component of the magnetic field has to decrease with height, so we consider $d_{3}\equiv0$. As a result the $z$-component of the magnetic field reduces to
 \begin{equation}
B_{z}=d_{1}J_{0}(\nu r)\cdot d_{4} e^{-\nu z}=d_{5}e^{-\nu
z}J_{0}\left(\nu r\right)=B_{0}e^{-\nu z}J_{0}\left(\nu r\right),
\label{45}
\end{equation}
where $B_{0}$ is the field strength at $r=z=0$.

Using the solenoidal condition, it is easy to show that the radial component of the magnetic field can be written as
\begin{equation}
B_r=B_0e^{-\nu z}J_1(\nu r).
\label{46}
\end{equation}

The particular form of the $z$-component of the magnetic field is
essential for determining the variation of the radius of the tube
with height. Assuming that the magnetic flux in the normal
direction to the cross-section of the tube is conserved (i.e.
${\bf A}\cdot {\bf B}=const$, where ${\bf A}=A{\bf n}$ with ${\bf
n}$ the unit vector perpendicular to the cross-section of the tube pointing upward) it is
easy to show that the variation of the radius of the tube is
described by
\begin{equation}
R(z)=R(0)e^{\nu z/2}\left[\frac{J_0(\nu R(0))}{J_0(\nu R(z))}\right]^{1/2},
\label{47}
\end{equation}
where $R(0)$ is the radius of the tube at $z=0$. The above transcendental equation is solved numerically using the Newton-Raphson method. The variation of the magnetic fields lines are shown in Figure 1.


\begin{figure}    
   \centerline{\includegraphics[width=0.6\textwidth,clip=]{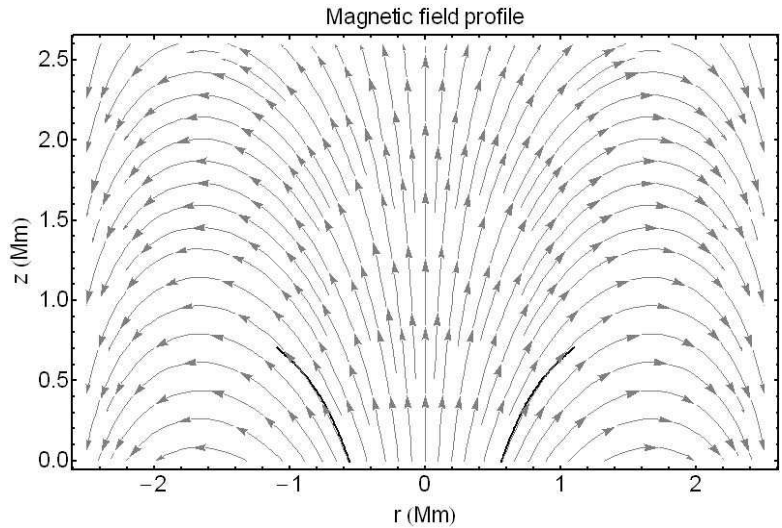}}
              \caption{Representation of the magnetic field lines and
              the flux tube whose boundary is calculated with the help of Eq. (\ref{47}). The longitudinal axis of the tube is situated at $R=0$.}
   \label{F-2}
   \end{figure}

The magnetic field lines diverge as initially assumed and the
inclination of the field lines become larger with the distance
from the axis of the tube. The axis of the flux tube is situated
at $R=0$ and the radius of the tube at $z=0$ is chosen to be 560
km while at the height of 750 km the radius becomes 1160 km. The
value of $\nu$ is arbitrary and its value $1.44\times 10^{-6}$ was
chosen in such a way to ensure that the decay of the magnetic
field with height is realistic. In this way, the magnetic field at
$z=0$ and $r=0$ is 250 G while at the height of 750 km this
becomes 85 G.

\begin{figure}    
   \centerline{\includegraphics[width=0.7\textwidth,clip=]{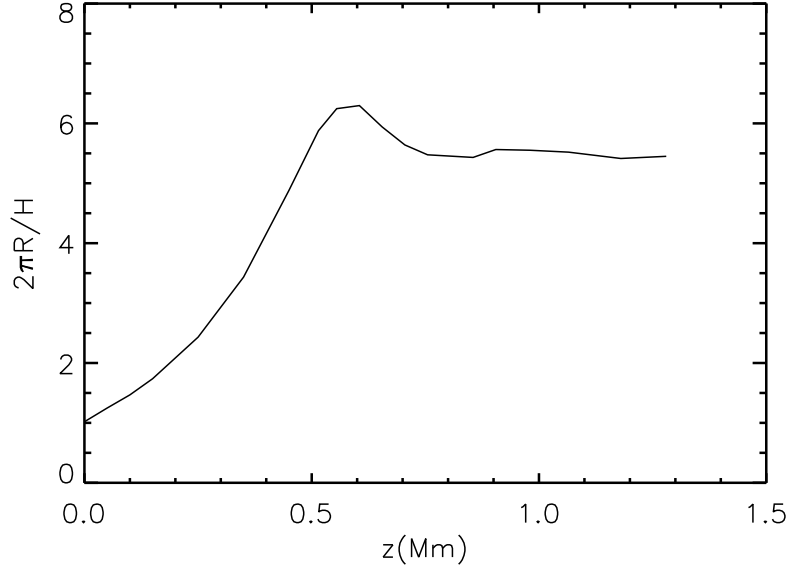}}
              \caption{The variation of $2\pi R/H$ in the solar photosphere. According to the standard criterion, a flux tube is classified as thin if this quantity is much less than one.}
   \label{F-2.2}
   \end{figure}

With the help of the tube's radius and the scale-height determined from the VAL-C model, we can finally justify our choice of thick flux tube in the solar photosphere (see Fig \ref{F-2.2}). As we specified earlier, a magnetic flux tube is thin provided $2\pi R/H\ll 1$. Combining our findings for the radius and a realistic atmospheric model it is obvious that the criterion for a thin flux tube is not satisfied for the whole height range we considered.
Now that the magnetic structure of the model is established we can study the dynamics of waves. Small amplitude disturbances of the equilibrium state of the gravitationally stratified atmosphere inside the tube can be described by the linear MHD equations in cylindrical coordinates
\begin{equation}  \label{1}
  \frac{\partial\rho^{'}}{\partial t}A+\rho\frac{\partial A^{'}}{\partial t}+\nabla\cdot\left(\rho A
  \textbf{v}\right)=0,
   \end{equation}
  \begin{equation}
\rho\frac{\partial\textbf{v}}{\partial t}=-\nabla
p^{'}+\frac{1}{\mu}\left[\left(\nabla\times\textbf{B}^{'}\right)\times\textbf{B}_{0}\right]+\textbf{g}\rho^{'},
\label{2}
\end{equation}
\begin{equation}
\frac{\partial p^{'}}{\partial t}+\gamma
p\left(\nabla\cdot\textbf{v}\right)+\left(\textbf{v}\cdot\nabla\right)p=0,
\label{3}
\end{equation}
\begin{equation}
\nabla\cdot\textbf{B}^{'}=0. \label{4}
\end{equation}
In the above equations, $p(z)$, $\rho(z)$,
$\textbf{B}_{0}=(B_{r},0,B_{z})$, $A(z)$
 denote the equilibrium pressure,
density, magnetic field, cross-section area of the tube. The
dashed quantities represent perturbations,
$\textbf{v}=(v_{r},0,v_{z})$ is the velocity perturbation, $\mu$
is the magnetic permeability of free space, \emph{g} is the
gravitational acceleration and $\gamma$ is the adiabatic index (from now one the dash will always denote a perturbation).

The above equations will be supplemented by the transverse pressure balance condition that must be satisfied at every height of the tube
\begin{equation}
p^{'}+\frac{B_{0}\cdot B^{'}}{2\mu}=P_{e}, \label{5}
\end{equation}
where $P_{e}$ is the total pressure in the exterior environment (includes kinetic and magnetic contributions but it's exact form is not important for the present analysis). In addition, we require that the magnetic flux along the tube is
constant, so
\begin{equation}
A(z) B^{'}_{z}+A^{'}(z)B_{z}=ct .\label{6}
\end{equation}

\section*{The evolutionary equation for slow MHD waves in thick flux tubes}

This section is devoted to the determination of the evolutionary equation describing the propagation of slow sausage waves in thick flux tubes under photospheric conditions. In order to obtain a physically acceptable equation some assumptions will be made regarding the direction of the dominant dynamics.

Using Eq. (\ref{1}), (\ref{3}), (\ref{4}), (\ref{5}), (\ref{6})
and the radial and longitudinal components of the momentum
equation (\ref{2}), it is rather straightforward to show that
these equations can be reduced to
\begin{equation}
\frac{\partial^{2}v_{r}}{\partial
t^{2}}=c_{0}^{2}\left(\frac{1}{r}\frac{\partial v_{r}}{\partial
r}-\frac{v_{r}}{r^{2}}+\frac{\partial^2 v_{r}}{\partial
r^{2}}\right)+\frac{\tau}{\rho}\frac{\partial p}{\partial
z}+\frac{B_{z}}{\mu\rho}\left(\frac{\partial^{2}
B^{'}_{r}}{\partial z\partial t}-\frac{\partial^{2}
B^{'}_{z}}{\partial r\partial t}\right), \label{7}
\end{equation}
and
\begin{eqnarray}
\frac{\partial^{2} v_{z}}{\partial
t^{2}}&=&\frac{1}{\rho}\frac{\partial}{\partial z}\left[\gamma
p\left(\frac{v_{r}}{r}+\frac{\partial v_{r}}{\partial
r}+\frac{\partial v_{z}}{\partial z}\right)+v_{z}\frac{\partial
p}{\partial
z}\right]-\frac{B_{r}}{\mu\rho}\left(\frac{\partial^{2}
B^{'}_{r}}{\partial z\partial t}-\frac{\partial^{2}
B^{'}_{z}}{\partial r\partial t}\right)  \nonumber \\
  && {}+g\left[\frac{1}{A\rho}\nabla\cdot\left(\rho
A\textbf{v}\right)-\frac{1}{B_{z}}\frac{\partial
B^{'}_{z}}{\partial t}\right],
    \label{8}
\end{eqnarray}
where we assumed that the variation of $v_{z}$ with respect to the coordinate
$r$ is constant and we write $\frac{\partial v_{z}}{\partial r}=\tau$. In the above equations $c_{0}$ is the sound speed defined as $c_{0}(z)=\sqrt{{\gamma
p(z)}/{\rho(z)}}$.

Equation (\ref{3}) and (\ref{5}) can be combined into
\begin{equation}
\frac{\partial B^{'}_{r}}{\partial t}=\left[\frac{\partial
P_{e}}{\partial t}+\gamma
p\left(\nabla\cdot\textbf{v}\right)+v_{z}\frac{\partial
p}{\partial z}- \frac{\partial B^{'}_{z}}{\partial
t}\right]\frac{\mu}{B_{r}}. \label{9}
\end{equation}
In a strong magnetic field slow waves propagate predominantly
along the field and especially in magnetic flux tubes the slow
mode is strongly wave guided along the tube. In his study Roberts (2006) found a simple method for extracting
information about slow modes from the MHD equations without the
need to calculate the behaviour of all the MHD modes, introducing a
spatial and temporal scaling of the form
\begin{equation}
r=\epsilon R, z=Z, t=T,
 \label{10}
\end{equation}
\begin{equation}
v_{r}=\epsilon u_{r}, v_{z}=u_{z}, B^{'}_{r}=\epsilon b^{'}_{r},
B^{'}_{z}=b^{'}_{z}.
 \label{11}
\end{equation}
For $\epsilon\ll 1$ the transverse components and derivatives are much larger than the longitudinal ones, so the slow mode related terms can be depicted as the ones containing the lowest order of $\epsilon$. Using this scaling we ensure that the dynamics of slow waves can be derived without the need of treating other waves and that the dominant dynamics of the plasma is corresponding to slow MHD waves.

Applying the above scaling to Eq. (\ref{9}) and collecting terms of the same order of $\epsilon$ we obtain (after returning to the initial notation)
\begin{equation}
\frac{B_r}{\mu}\frac{\partial B^{'}_{r}}{\partial
t}=\frac{\partial P_{e}}{\partial t}, \label{12}
\end{equation}
and
\begin{equation}
\frac{B_z}{\mu}\frac{\partial B^{'}_{z}}{\partial t}=\gamma
p\left(\nabla\cdot\textbf{v}\right)+v_{z}\frac{\partial
p}{\partial z}.\label{13}
\end{equation}
Applying the same scaling to Eg.(\ref{7}) we obtain
\begin{equation}
\frac{\partial v_{r}}{\partial t}=\frac{B_{z}}{\mu
B^{2}_{r}}\left(\frac{\partial P_{e}}{\partial
z}B_{r}-\frac{\partial B_{r}}{\partial z}P_{e}\right). \label{14}
\end{equation}

Finally, using Eqs. (\ref{8}), (\ref{12}) and (\ref{13}) one can derive the
expression
\begin{equation}
\frac{\partial^{2} v_{z}}{\partial
t^{2}}-c_{o}^{2}\frac{\partial^{2} v_{z}}{\partial
z^{2}}-\frac{\partial v_{z}}{\partial
z}\left(\frac{1}{\rho}\frac{\partial p}{\partial
z}+\Gamma\right)-v_{z}\Lambda=c^{2}_{0}\frac{\partial
\Xi}{\partial z}, \label{15}
\end{equation}
where we introduced the notations
\begin{equation}
\Gamma=\frac{\gamma}{\rho}\frac{\partial p}{\partial
z}+g-\frac{c^{2}_{0}}{v^{2}_{Az}}g, \quad  v^{2}_{Az}=\frac{B_{z}^{2}}{\mu\rho}, \label{16}
\end{equation}
\begin{equation}
\Lambda=\frac{1}{\rho}\frac{\partial^{2} p}{\partial
z^{2}}+\frac{g}{A}\frac{\partial A}{\partial
z}+\frac{g}{\rho}\frac{\partial\rho}{\partial
z}-\frac{g}{\rho}\frac{\partial p}{\partial z}\frac{1}{v^{2}_{Az}},
\label{18}
\end{equation}
\begin{equation}
\Xi=\frac{1}{r}\frac{\partial}{\partial r}\left[\frac{rB_{z}}{\mu
B_{r}^{2}}\left(\frac{\partial P_{e}}{\partial
z}B_{r}-\frac{\partial B_{r}}{\partial z}P_{e}\right)\right].
\label{19}
\end{equation}
Equation (\ref{15}) is a partial differential equation describing the evolution of the slow wave variable, $v_{z}$, in space and time while the term on the right hand side depends solely on external pressure and the values of the equilibrium magnetic field.

Assuming a solution of the form $v_{z}=Qe^{\lambda(z)z}=Qe^{c(z)}$
and after dividing Eq. (\ref{15}) by $e^{c(z)}$, we arrive at
\begin{eqnarray}
 \lefteqn{\frac{\partial^{2} Q}{\partial t^{2}}-c_{0}^{2}\frac{\partial^{2}
Q}{\partial z^{2}}-\frac{\partial Q}{\partial
z}\left(c^{2}_{o}2C+\frac{1}{\rho}\frac{\partial p}{\partial
z}+\Gamma\right) } \nonumber \\
   && -Q\left[c_{0}^{2}\left(C^{2}+C^{'}\right)+\left(\frac{1}{\rho}\frac{\partial
p}{\partial
z}+\Gamma\right)C+\Lambda\right]=c_{0}^{2}\frac{\partial
\Xi}{\partial z}\cdot e^{-c(z)},
    \label{20}
\end{eqnarray}
where
\begin{equation}
C(z)=\frac{d c(z)}{dz}, C(z)^{'}=\frac{d C(z)}{dz}.\label{21}
\end{equation}

Next, we are going to choose the function $C(z)$ so that the coefficient of ${\partial Q}/{\partial z}$ vanishes. As a result, we obtain that the dynamics of the slow sausage waves in thick flux tubes is given by a Klein-Gordon equation
\begin{equation}
\frac{\partial^{2} Q}{\partial t^{2}}-c_{0}^{2}\frac{\partial^{2}
Q}{\partial z^{2}}+Q\omega_{c}^{2}=c_{0}^{2}\frac{\partial
\Xi}{\partial z} e^{-c(z)}, \label{22}
\end{equation}
where the term multiplying \emph{Q} is the cut-off frequency and
its expression is given by
\begin{equation}
\omega_{c}^{2}=\frac{1}{4c_0^2}\left(\frac{1}{\rho}\frac{\partial
p}{\partial
z}+\Gamma\right)^2+c^{2}_{0}\frac{\partial }{\partial
z}\left(\frac{\frac{1}{\rho}\frac{\partial p}{\partial
z}+\Gamma}{2c^{2}_{0}}\right)-\Lambda. \label{24}
\end{equation}

The cut-off frequency has a complex form and it contains all the
main plasma parameters, being dependent on pressure, density,
cross section of the tube and their variation with height, as well
as the on the plasma beta parameter and the chosen profile of the
magnetic field. Eq. (\ref{22}) shows that slow waves will
propagate with the internal sound speed in accordance with the
dispersion diagram determined by Edwin83 for photospheric
conditions. This is contrast to the findings of wave propagation
in thin flux tubes, where the propagation speed is the tube speed.
Since at this stage we are not interested in the eigenfunction of
Eq. (\ref{22}) we will not solve this equation, instead we
concentrate on the variation of the coefficient given by Eq.
(\ref{24}) with height and radial distance.

In principle the radial component of the magnetic field would considerably modify the value of the Brunt-$V\ddot{a}is\ddot{a}l\ddot{a}$ buoyancy frequency but since our magnetic field was assumed force-free, the magnetic contribution to this important quantity vanishes and its value is given by the classical formula.

\section*{Results and Discussions }

Having determined the analytical form of the cut-off frequency
(see Eq. (\ref{24})), we are now interested in its profile inside
the expanding tube and its variation with height and radial
distance. In order to obtain actual values for this important
physical parameter we consider the observational data provided by
the VAL III C model Vernazza1981 for the quiet sun (consider that no
extreme magnetic phenomena are taking place in the proximity of
the flux tube) because of its simplicity.

\begin{figure}    
   \centerline{\includegraphics[width=0.7\textwidth,clip=]{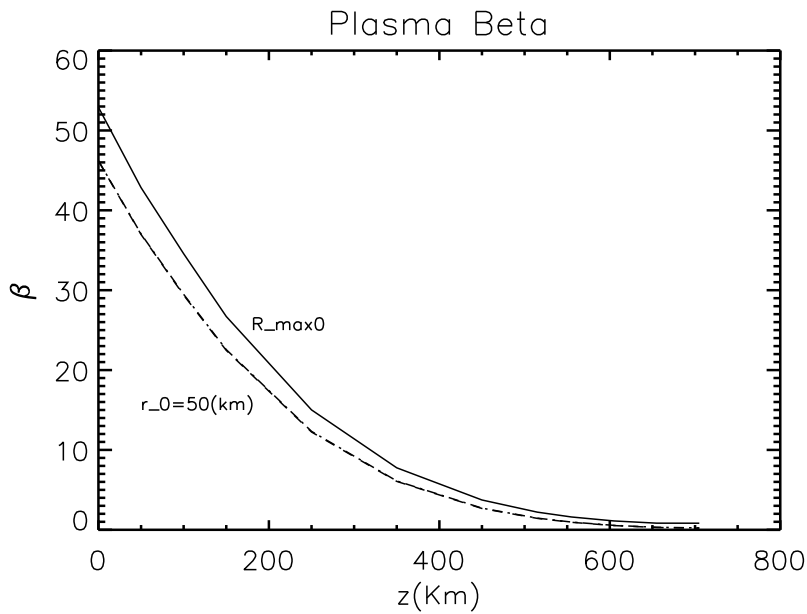}}
              \caption{The variation of plasma-beta with height for two values of he radial distance from the longitudinal axis of the tube (at $r=0$). }
   \label{F-3.1}
   \end{figure}

The magnetic field profile derived earlier allows us to find the
value of its components in every point inside the tube. The value
of the magnetic field is important for the present analysis as
this will control the evolution of other physical parameters
through the total pressure balance. The VAL III C model data
provided us with values for pressure, density, and temperature at
certain heights. We assume that the pressure extracted from the
atmospheric model represents the kinetic pressure in the external
region. We consider that the tube is embedded in a quiescent
environment, the magnetic field strength at the footpoint inside
the tube was taken to be 250 G. Because of the poor fit of the
magnetic field profile with the observational data, we were
constrained to consider additionally a very weak magnetic field in
the surrounding environment (60 G). Since the exact value of the magnetic
field profile of the tube can be determined with the help of Eqs.
(\ref{45})-(\ref{46}) we can derive the equilibrium pressure in
the interior of the tube at different heights. For simplicity we
assumed that in the radial direction the magnetic flux tube is
isothermal resulting in an equilibrium pressure that depends only
on height. Once the equilibrium pressure was determined, the
hydrostatic pressure balance was used to determine the profile of
the density inside the tube. With the help of these values, the
determination of characteristic speeds (sound speed and Alfv\'en
speed) was straightforward. These speeds can help us determining
the plasma-beta ($\beta$) inside the tube. Through the Alfv\'en
speed, the plasma-beta will depend not only on height but also on
radial direction. The variation of $\beta$ for two different
radial distances is plotted in Fig. (\ref{F-3.1}) where $R_{max0}$
represents the maximum radius of the tube determined by means of
Eq. (\ref{47}) and a second level at which the plasma-beta is
calculated was taken at $R_0= 50 km$ from the axis of the tube.
The distance of this reference level from the axis of the tube is
also changing with height.
 It is obvious that the value of $\beta$ decays with height and for a given height it
 becomes smaller closer to the axis of the tube. The results are
 in concordance with the recent realistic plasma beta profiles
(ssFedunVerth2011) for heights above 500 km. This difference it is likely to be due
 to the force field assumption that we have chosen for our
 theoretical model.

\begin{figure}    
   \centerline{\includegraphics[width=0.7\textwidth,clip=]{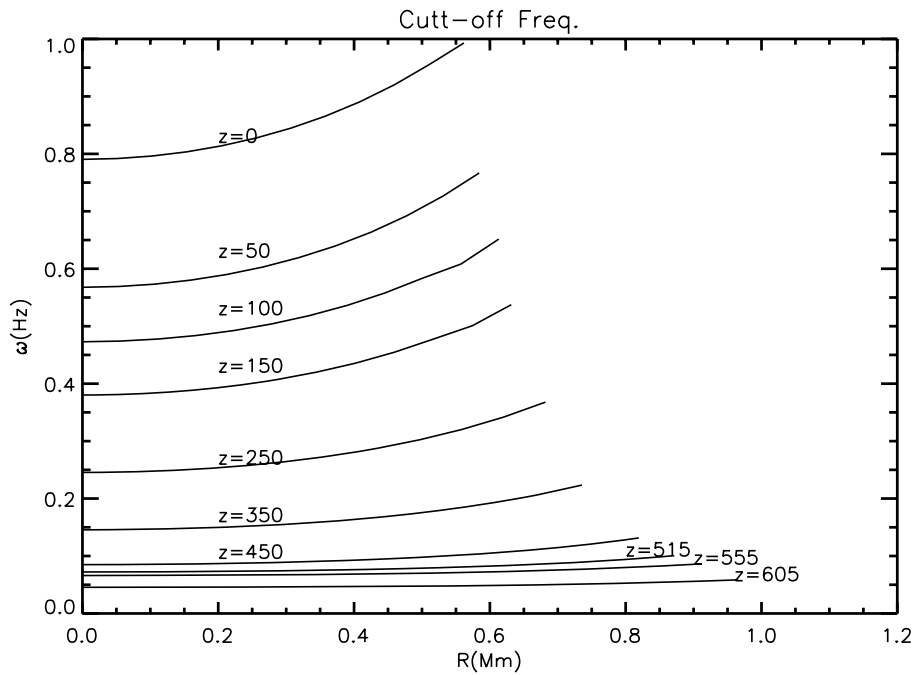}}
              \caption{Cut-off frequency profile (height and radial dependence) inside the
              tube (up to 605 km in height).}
   \label{F-3}
   \end{figure}

Now that all physical quantities entering the expression of the cut-off frequency are determined, we can plot the dependence of the cut-off frequency on the two directions (see Fig.\ref{F-3}).
In Fig.\ref{F-3} we plot the obtained values for the cut-off frequency with respect to the position along the tube radius for various heights. The values of the cut-off frequency are decreasing with height, meaning that waves can propagate easier if the driver of the disturbance is situated above the foot point, inside the loop, and if the waves are propagating closer to the axis rather than closer to the tube's boundary. Due to the radial dependence of the cut-off frequency, we can conclude that the region near the axis of the tube allows a wider range of wave frequencies to propagate upwards that the region near the boundary. Our results show that a magnetic flux tube acts as a frequency filter for upward propagating waves.
\begin{figure}
   \centerline{\includegraphics[width=0.8\textwidth,clip=]{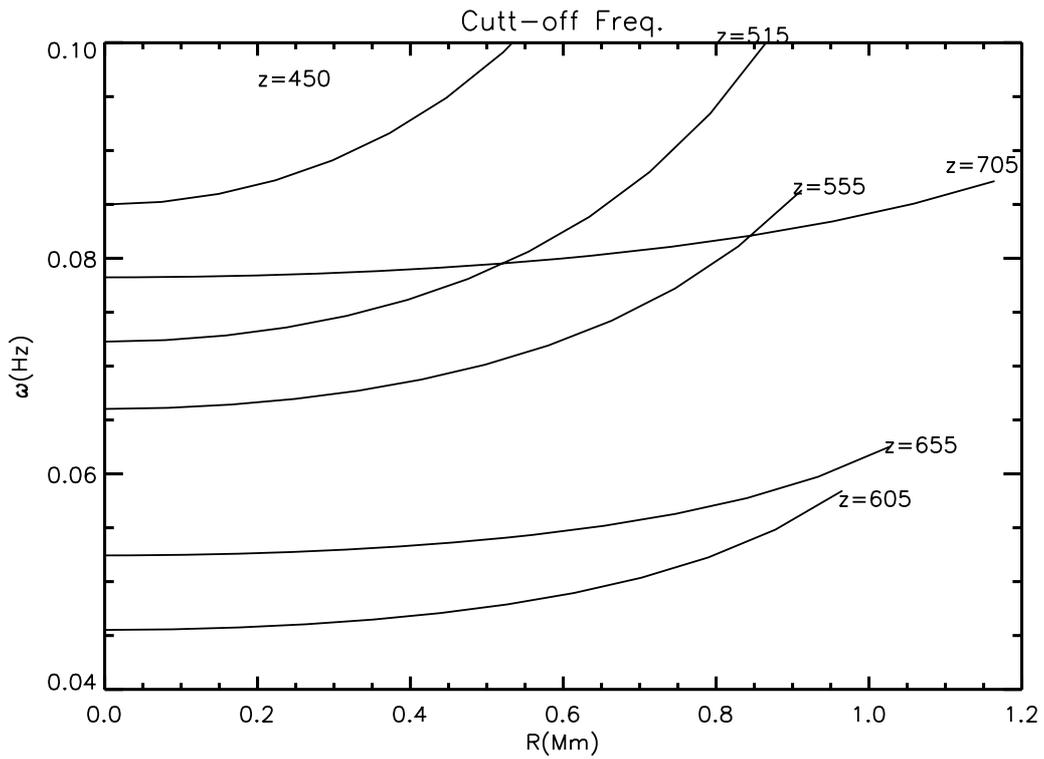} }
              \caption{Cut-off frequency profile(height and radial dependence) inside the
              tube (from 450km to 705km in height)}
   \label{F-4}
   \end{figure}
A "blown-up" section of Fig. \ref{F-3} for small values of the
cut-off frequency is shown in Fig. \ref{F-4} and the pattern of
the variation of this frequency with height and radial dependence
is maintained.

\begin{figure}    
   \centerline{\includegraphics[width=0.8\textwidth,clip=]{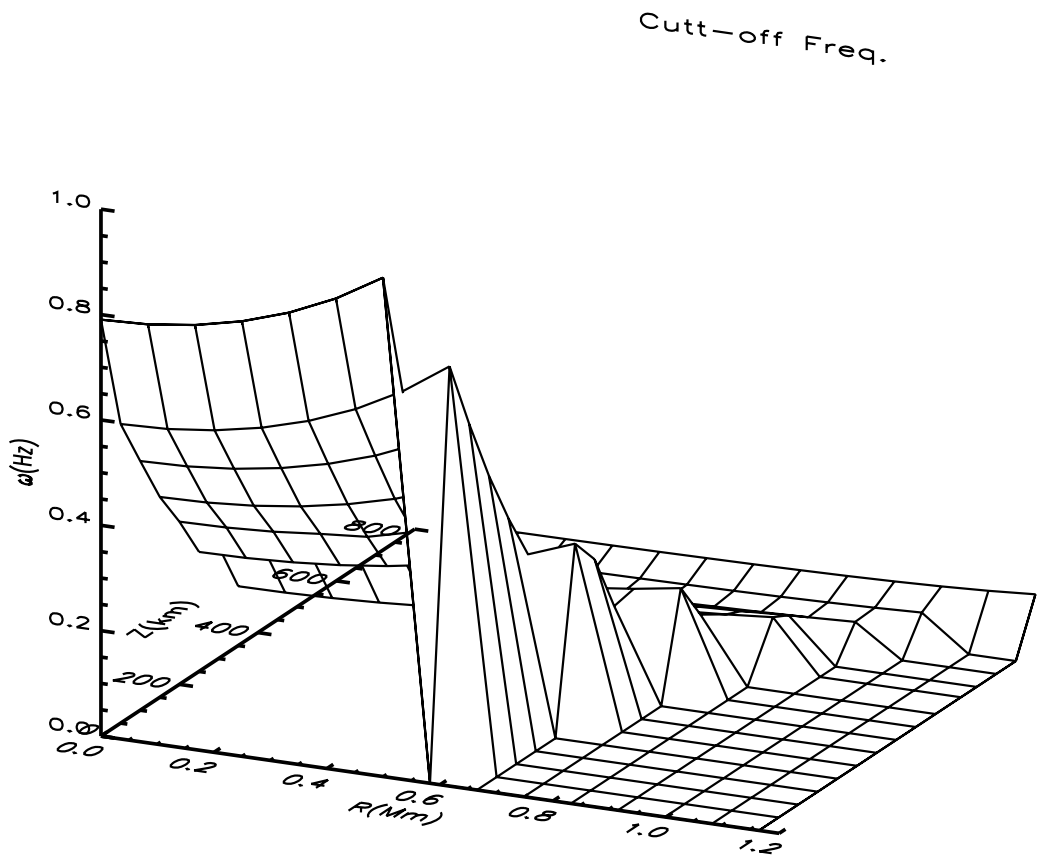} }
              \caption{The variation of the cut-off frequency with height and radial distance from the axis of the flux tube. }
   \label{F-5}
   \end{figure}

The 3D dependence of the cut-off frequency with the two coordinates can be seen in Fig \ref{F-5} where the edge of the tube can be seen as the point in the ($r-z$) plane where the cut-off frequency drops to zero.

The recovered cut-off frequencies are somehow larger than the
expected values, especially for lower heights. Although our
intention was to show how a magnetic flux tube can act as a
frequency filter for slow sausage waves propagating in the lower
photosphere, the recovered frequencies are not in line with the
expected values indicating that some important physical effect(s)
was neglected. One possibility that can reduce the value of the
cut-off frequency is the inclusion of dissipative effects. In an
earlier paper Ballai2006 studied the propagation of slow
MHD waves in a stratified and viscous plasma and they found that
the evolutionary equation of these waves was described by the
Klein-Gordon-Burgers equation where the cut-off frequency was
reduced by a term proportional to the square of the coefficient of
viscosity. It is possible that if realistic transport mechanisms
are taken into account (including ambipolar diffusion due to the
partial ionisation of photospheric plasma) the true value of the
cut-off frequency can be reduced to an acceptable level. Another
important factor that could influence the value of the cut-off
frequency is the consideration of a filed-aligned equilibrium
flow, something that has been observed in the solar photosphere.
Finally, discrepancies in the value of the cut-off frequency can
appear due to the simplistic model our paper used. An improved
model where all restrictions imposed in our paper are relaxed
would require a fully numerical investigation.

\section*{Conclusion} 

In this study we investigated the variation of the cut-off frequency imposed by the plasma in the case of propagation of slow sausage waves in thick flux tubes in the solar atmosphere. The prescribed magnetic field is force-free. The particular choice of the magnetic filed allowed us to derive an evolutionary equation of waves in the form of a Klein-Gordon equation. Apart from describing the dynamics of slow waves, this equation also gives the value of the cut-off frequency, i.e. the lower limit of frequencies at which waves can still propagate through the magnetic and stratified plasma. The total pressure balance equation satisfied at every height of the tube was used to connect the values of the VAL III C model atmospheric model with the plasma inside the flux tube.

Our intention was not to solve the evolutionary equation; given the built-up of our model, this task would require an extensive numerical analysis. Instead we focussed our attention to one of the coefficients of the evolutionary equation that represents the cut-off frequency. Our derived values for the cut-off frequency show that the tube acts as a frequency filter, the cut-off values depending on the radial distance and height. The recovered values are somehow larger than the expected, especially at lower heights but this could be attributed to either the simplistic model employed by our study or to some essential physics that was neglected.

\section{Acknowledgements}

The present study was completed while one of us (AP) visited the Solar Physics and Space Plasma Research Centre (SP2RC) at the University of Sheffield. IB was financially supported by NFS Hungary (OTKA, K83133).

\section*{References}

Asai01: Asai,~A. et al.: 2001, \textit{Periodic acceleration of elecrons
in the 1998 November 10 solar flare}, ApJ, \textbf{562},
L103-L106.

Aschwanden99: Aschwanden,~M.J., Fletcher,~L., Schrijver,~C.J. and Alexander,~D.:
1999, \textit{Coronal loop oscillations observed with the
Transition Region and Coronal Explorer}, ApJ, \textbf{520},
880-894.

Aschwanden2006: Aschwanden,~M.J.: 2006, \textit{Physics of the Solar Corona},
Printer and Praxis Publishing, Chichester, UK.

Ballai2006: Ballai,~I., Erd\'elyi,~R. and Hargreaves,~J: 2006, \textit{Slow
magnetohydrodynamic waves in stratified and viscous plasmas},
Phys.Plasmas, \textbf{13}.

Ballai2011: Ballai,~I., Jess,~D.B. and Douglas,~M.: 2011, \textit{TRACE
observations of the driven loop oscillations}, AAp,
\textbf{534}.

Berghmans99: Berghmans,~D., Clette,~F.: 1999, \textit{Active region EUV
transient brightenings- First Results by EIT of SOHO JOP 80},
SolPhys, \textbf{186}, 207-229.

Berghmans01: Berghmans,~D., McKenzie,~D. and Clette,~F.: 1999, \textit{Active
region transient brightenings. A aimultaneous view by SXT, EIT and
TRACE}, AAp, \textbf{369}, 291-304.

DeForest98: DeForest,~C.E. and Gurman,~J.B.: 1998, \textit{Observation of
quasi-periodic compressive waves in solar polar plumes}, ApJ,
\textbf{501}, L217-L220.

De Moortel00: De Moortel,~I. and Hood,~A.W.: 2000, \textit{Wavelet analysis and
the determination of coronal plasma properties}, AAp,
\textbf{363}, 269-278.

Dorotovic08: Dorotovic,~I., Erd\'elyi,~R. and Karlovsky,~V.: 2008, IAU Symp.(ed),
\textit{Identification of linear slow sausage waves in magnetic
pores}, \textbf{247}.

Dorotovic12: Dorotovic,~I., Erd\'elyi,~R., Freij,~N., Karlovsky,~V., Marquez,~I.:
2012, \textit{On standing sausage waves in photospheric magnetic
waveguides}, AAp, \textbf{1210}.

Edwin83: Edwin,~P.M. and Roberts,~B.: 1983, \textit{Wave propagation in a
magnetic cylinder}, SolPhys, \textbf{99}.

Erd108: Erd\'elyi,~R. and Hargreaves,~J.: 2008, \textit{Wave propagation in
steady stratified one-dimentional cylindrical waveguides}, AAp,
\textbf{483}, 285-295.

Erd08: Erd\'elyi,~R. and Taroyan,~Y.: 2008, \textit{Hinode EUV
spectroscopic observations of coronal oscillations}, AAp,
\textbf{489}, L49-L52.

Fedun2011: Fedun,~V. et al.: 2011, \textit{Numerical Modeling of
Footpoint-Driven Magneto-Acoustic Wave Propagation in a  Localized
Solar Flux Tube}, ApJ, \textbf{727}, 17-31.

FedunVerth2011: Fedun,~V., Verth,~G., Jess,~D.B., and Erd\'elyi,~R.: 2011,
\textit{Frequency filtering of torsional Alfven waves by
chromospheric magnetic filed}, ApJ, \textbf{740}.

Fujimura09: Fujimura,~D. and Tsuneta,~S.: 2009, \textit{Properties of
magnetohydrodynamic waves in the solar photosphere obtained with
HINODE}, ApJ, \textbf{702}, 1443-1457.

Gilbert04: Gilbert,~H.R. and Holzer,~T.E.: 2004, \textit{Chromospheric waves
observed in the He I spectral line ($\lambda=10830 \AA$): a closer
look}, ApJ, \textbf{610}, 572-587.

Katsiyannis03: Katsiyannis,~A.C., Williams,~D.R., McAteer,~R.T.J.,
Gallagher,~P.T., Keenan,~F.P., Murtagh,~F.: 2003, \textit{Eclipsa
observations of high-frequency oscillations in active region
coronal loops}, AAp, \textbf{406}, 709-714.

Marsh09: Marsh,~M.S., Walsh~R.W. and Plunkett,~S.: 2009,
\textit{Three-dimensional coronal slow modes: towards
three-dimensional seismology}, ApJ, \textbf{697}, 1674-1680.

McEwan06: McEwan,~M.P. and De Moortel,~I.: 2006, \textit{Longitudinal
intensity oscillations observed with TRACE: evidence of fine-scale
structure}, AAp, \textbf{448}, 763-770.

Nightingale99a: Nightingale,~R.W., Aschwanden,~M.J., Hurlburt,~N.E.: 1999,
\textit{Time Variability of EUV Brightning in Coronal Loops
Observed with TRACE}, SolPhys, \textbf{190}, 249-265.

Nightingale99b: Nightingale,~R.W., Aschwanden,~M.J., Hurlburt,~N.E.: 1999,
\textit{Time Variability of Coronal Loops observed by TRACE},
American Astron. Society, \textbf{31}, 961.

Ofman2000: Ofman,~L.: 2000, ASP Series (ed.)\textit{Propagation and
Dissipation of Slow Magnetosonic Waves in Coronal Plumes},
University of Chicago Press (USA), \textbf{205}.

Ofman2008: Ofman,~L. and Wang,~T.J.: 2008, \textit{Hinode observations of
transversal waves with flows in coronal loops}, AAp,
\textbf{482}, L9-L12.

RaeRoberts1982: Rae,~I.C., Roberts,~B.: 1982, \textit{Pulse Propagation in a
Magnetic Flux Tube}, ApJ, \textbf{256}, 761-767.

Roberts2001: Roberts,~B.: 2001, entry \textit{Solar Photospheric Magnetic Flux
Tubes: Theory}, \textit{Encyclopedia of Astronomy and
Astrophysics}, eds Paul Murdin, Nature Publishing Group.

Roberts2006: Roberts,~B.: 2006, \textit{Slow magnetohydrodynamic waves in the
solar atmosphere},  Philos. Trans. R. Soc. London,
\textbf{634}, 447-460.

Schrijver1999: Schrijver,~C.J., Hurlburt,~N.E., Engvold,~O., Harvey,~J.W.: 1999,
\textit{Physics of the solar corona and the transition region.
Part 1,2. Proceedings. Workshop, Monterey, CA (USA), SolPhys},
\textbf{190, 193}, 1-497, 1-297.

Vernazza1981: Vernazza,~J.E., Avrett,~E.H., Loeser,~R.: 1981, \textit{Structure
of the Solar Chromosphere III. Models of the EUV Brightness
Components of the Quiet Sun},ApJ, \textbf{45}, 635-725.

Wang03: Wang,~T.J., Solanski,~S.K., Curdt,~W., Innes,~D.E.,
Dammasch,~I.E., Kliem,~B.: 2003, \textit{Hot coronal loop
oscillations observed with SUMER: Exemples and statistics},
AAp, \textbf{406}, 1105-1121.

Williams01: Williams,~D.R. et al.: 2001, \textit{High-frequency oscillations
in a solar active region coronal loop}, mnras, \textbf{326},
428-436.

Williams02: Williams,~D.R. et al.: 2002, \textit{An observational study of a
magneto-acoustic wave in the solar corona}, mnras,
\textbf{336}, 747-752.

Zirin72: Zirin,~H. and Stein,~A.: 1981, \textit{Observations of running
penumbral waves}, ApJs, \textbf{178}, L85-L87.

\end{document}